\documentclass[twocolumn,showpacs,preprintnumbers,amsmath,amssymb,prb,floatfix]{revtex4-1}
\usepackage[bbgreekl]{mathbbol}
\usepackage{graphicx}
\usepackage{dcolumn}
\usepackage{bm}
\usepackage{mathtools}
\usepackage{hyperref}
\usepackage{color}
\usepackage[a4paper]{anysize}
\graphicspath{{fig/}{img/}}
\usepackage{placeins}

\usepackage[exponent-product=\cdot , decimalsymbol=. ,per-mode=fraction]{siunitx}
\usepackage{braket}
\usepackage{multirow}
\usepackage[justification=centerlast,singlelinecheck=false]{caption} 
\usepackage[label font=bf,labelformat=simple,justification=raggedright, margin = 0 pt, skip = -11pt]{subcaption}

\def\gsim{\lower.35em\hbox{$\stackrel{\textstyle>}{\textstyle\sim}$}}
\def\lsim{\lower.35em\hbox{$\stackrel{\textstyle<}{\textstyle\sim}$}}

\begin{document}
\title{Flat-band ferromagnetism in twisted bilayer graphene}

\author{R. Pons}
\affiliation{Institut f\"ur Theoretische Physik, Philosophenweg 19, Universit\"at Heidelberg, Germany}
\author{A. Mielke}
\affiliation{Institut f\"ur Theoretische Physik, Philosophenweg 19, Universit\"at Heidelberg, Germany}
\author{T. Stauber}
\affiliation{Instituto  de Ciencias de Materiales de Madrid, CSIC, E-28049, Madrid, Spain}
\affiliation{Institut f\"ur Theoretische Physik, Universit\"at Regensburg, Germany}
\begin{abstract}
We discuss twisted bilayer graphene (TBG) based on a theorem of flat band ferromagnetism put forward by Mielke and Tasaki. According to this theorem, ferromagnetism occurs if the single particle density matrix of the flat band states is irreducible and we argue that this result can be applied to the quasi-flat bands of TBG that emerge around the charge-neutrality point for twist angles around the magic angle $\theta\sim1.05^\circ$. We show that the density matrix is irreducible in this case, thus predicting a ferromagnetic ground state for neutral TBG ($n=0$). We then show that the theorem can also be applied only to the flat conduction or valence bands, if the substrate induces a single-particle gap at charge neutrality. Also in this case,  the corresponding density matrix turns out to be irreducible, leading to ferromagnetism at half filling ($n=\pm2$).
\end{abstract}
\maketitle
\section{Introduction} 
Twisted bilayer graphene (TBG) has attracted much attention due to the recent discovery of superconductivity.\cite{Cao18b,Yankowitz19,Moriyama19,Codecido19,Shen19,Lu19,Chen19,Xu18,Volovik18,Yuan18,Po18,Roy18,Guo18,Dodaro18,Baskaran18,Liu18,Slagle18,Peltonen18,Kennes18,Koshino18,Kang18,Isobe18,You18,Wu18b,Zhang18,Gonzalez19,Ochi18,Thomson18,Carr18,Guinea18,Zou18,Gonzalez20,GonzalezSB20,StauberC20} 
Also correlated gaps were observed\cite{Kim17,Cao18a} that cannot be explained by a one-particle band-theory.\cite{Suarez10,Bistritzer11} The fact that interactions severely change the one-particle band structure has further been demonstrated in recent local probe experiments.\cite{Kerelsky19,Xie19,Jiang19,Choi19}

Also the emergence of flat-band ferromagnetism in intrinsic twisted bilayer graphene was predicted using first principle DFT-calculations.\cite{Yndurain19,Lopez-Bezanilla19} In fact, ferromagnetism seems to be present at all integer filling factors of the flat bands,\cite{Repellin19} and close to a van Hove singularity it was observed by local probe microscopy.\cite{LiuQiao19} Let us also note that based on maximally localized superlattice Wannier wave functions,\cite{Kang18,Koshino18,Po18x} an effective spin model suggests that the system is described by a ferromagnetic Mott insulator at quarter filling ($n=1$)\cite{Seo19} and half filling ($n=2$).\cite{Kang19} 

Even yet another kind of ferromagnetism can arise in the presence of topological bands that emerge due to a single-particle gap at charge neutrality. It is well-known that single-layer epitaxial graphene can develop a substrate-induced mass term,\cite{Zhou07} and if the TBG-sample is crystallographically aligned with respect to the underlying boron-nitride (BN) substrate, the adjacent graphene layer displays a gap due to the proximity effect.\cite{Hunt13} Considering only one valley, this induces a gap exclusively at one $K$-point in the Moir\'e Brillouin zone for large twist angles. But for small twist angles, the valence and conduction bands become completely gapped due to the enhanced interlayer coupling. The flat bands thus become Chern bands which leads to anomalous Hall ferromagnetism at filling factor $n=3$.\cite{Sharpe19,Serlin20}
This makes TBG and also related systems such as ABC-trilayer gaphene on a misaligned BN-substrate\cite{Chen19} an ideal platform to study the interplay between correlations and topology.   

The anomalous Hall ferromagnetism, a new state of matter, is characterised by a spin and valley-polarised ground state\cite{Bultinck19,ZhangMao19} and recent magnetoresistance measurements\cite{Serlin20} show non-monotonic behaviour consistent with skyrmion excitations.\cite{Chatterjee19} Hysteresis behaviour is further expected in non-linear photo-conductivities as they are proportional to the orbital magnetisation of the system.\cite{Liu20} And a Schwinger boson analysis with complementary density matrix renormalisation also predicts ferromagnetism at quarter and three-quarter filling, i.e., $n=1$ and $n=3$,\cite{WuKeselman19} which is also the conclusion of Ref. \onlinecite{Alavirad20} who analyze the ferromagnetic instability in terms of spin-density waves.

In this paper, we will discuss ferromagnetism using a general theorem initially put forward by Mielke\cite{Mielke93,Mielke1999} who showed that in a flat band at half filling there is a unique ferromagnetic ground state up to spin-degeneracy if and only if the density matrix of the single particle states forming the flat band is irreducible. A careful and readable proof of this theorem can be found in the book by Tasaki\cite{Tasaki2020}. We will show that this theorem can be applied to the 4 bands around charge neutrality in the case of pristine TBG. In the presence of a substrate induced gap, we will argue that it can also be exclusively applied to the two highest valence or two lowest conduction bands. In both cases, the resulting density matrix turns out to be irreducible, thus predicting ferromagnetism at the neutrality point ($n=0$) and at half-filling ($n=\pm2$), respectively. Let us finally mention  that  orbital effects\cite{Bahamon20,BultinckX20} are not included in our approach.

\section{Previous results} 
Before we discuss ferromagnetism in TBG, let us recall basic theorems and results concerning magnetic ground-states of graphene and related systems.
\subsection{Antiferromagnetism}
For single-layer graphene at half-filling, antiferromagnetism is stable beyond a critical Hubbard interaction $U\sim3.7$eV.\cite{Assaad13} Still, antiferromagnetism does normally not occur in flat bands, only ferrimagnetism.\cite{Lieb89} Nevertheless, triangular antiferromagnetism on the honeycomb lattice was predicted in the presence of a spin density wave lying on the bonds.\cite{Thomson18}

For antiferromagnetism or ferrimagnetism, one usually needs a bipartite lattice. For bipartite lattices, there can be a flat band at zero energy and if this is the case, one ends up with a ferrimagnet. 

\subsection{Flat-band ferromagnetism}
\label{Ferromagnetism}
Let us now summarize some general results for flat-band ferromagnetism to which we refer in this paper. A Hubbard model on an arbitrary lattice with a flat band at the bottom of the spectrum has ferromagnetic ground states if the band is at most half filled. At half filling, the ferromagnetic ground state is unique up to a $SU(2)$-spin-degeneracy if and only if the single particle density matrix formed by the degenerate single particle ground states is irreducible.\cite{Mielke93,Mielke1999} This result also applies to the case of a flat band at the top of the spectrum via particle-hole transformation and even extends to the case where the flat band lies somewhere in the spectrum using a perturbative argument.\cite{Mielke93} The perturbative argument is only valid for small Hubbard $U$, however, there is a (yet unproven) conjecture that the expectation value of $S^2$ in the ground state can only increase monotonically with $U$.\cite{Suto91} If this was true, the system would be ferromagnetic, independent of the Hubbard-$U$. 

For an almost, but not completely flat band, it has been proven for several classes of lattices that the ferromagnetism remains stable for sufficiently large $U$ if there is a gap between the flat band and the rest of the spectrum, see e.g. Refs. \onlinecite{Tasaki94,Tasaki96}. For a modified Kagome lattice, this is also true even though there is no gap \cite{Tanaka2003}. If we assume that this holds for TBG as well, which has a gap, we need to show that the single particle density matrix formed by the single particle states of the flat or almost flat bands in TBG is irreducible to obtain ferromagnetism. 

\section{Theoretical Approach}
\subsection{Application to TBG}
We will argue that the ground state of magic angle TBG is ferromagnetic by applying the findings for flat band ferromagnetism. We use the following theorem from Ref. \cite{Mielke1999} that we will state here again:\\

{\it The ferromagnetic ground state of the Hubbard model with $N_d$ sites and $N_e=N_d$ electrons is the unique ground state (up to the spin degeneracy due to the $SU(2)$ symmetry) if and only if the single-particle density matrix $\rho_{ij}$ is irreducible.}\\

The main quantity of our discussion is thus given by $\rho_{ij}$ and our analysis is divided into two steps: (i) First, to numerically calculate $\rho_{ij}$ for a given model, and (ii) second, to probe the resulting density matrix with respect to its irreducibility. 

However, we have not yet specified the underlying Hilbert space on which the density matrix is defined. Primarily, we are interested in discussing ferromagnetism at the neutrality point, and the Hilbert space is given by the four bands around the neutrality point, i.e., the two highest valence and two lowest conduction bands where both valleys are included. If the four bands are now separated from the remote bands by a  large enough single-particle gap, we can apply the above theorem as outlined in Sec. \ref{Ferromagnetism} - at least perturbatively.\cite{Mielke93}

We can also apply our analysis to discuss ferromagnetism at half-filling of the two lowest conduction {\it or} two highest valence band ($n=\pm2$). The density matrix is then defined only with respect to the two upper or the two lower bands. However, the conduction and valence bands must be separated by a large enough gap at the Dirac point that can be induced by a crystallographically aligned substrate.

\subsection{Models for TBG} 
We will consider two microscopic models to describe twisted bilayer graphene: (i) the continuum model (CM) first introduced by Lopes dos Santos, Peres, and Castro-Neto\cite{Lopes07,Mele10,Bistritzer11,Moon12} and (ii) the tight-binding model (TBM).\cite{Suarez10,Trambly10} For better comparison, we will only discuss twist angles corresponding to commensurate systems that can be characterized by the integer $i$. The twist angles are then given by $\cos\theta_i=\frac{3i^2+3i+0.5}{3i^2+3i+1}$. 
\subsubsection{Continuum model}
Representing twisted bilayer graphene in a plane-wave basis leads to the so-called continuum model. Assuming a symmetric interlayer coupling does not lead to a single-particle gap that separates the flat bands from the remote bands; still, a gap opens up by introducing an out-of-plane lattice relaxation to the sample. The corrugation can be modeled by an asymmetric interlayer coupling for the AA-stacked and AB-stacked regions, respectively, and Koshino et al. \cite{Koshino18} obtain the parameters $u = \SI{0.0797}{\electronvolt}$ and $u^{\prime} = \SI{0.0975}{\electronvolt}$. For a better comparison  to previous results, we prefer to use scaled parameters, i.e., $u = \SI{0.0898}{\electronvolt}$ and $u^{\prime} = \SI{0.11}{\electronvolt}$, thus fixing the interlayer coupling in the (isolated) AB-stacked region to $t = \SI{2.78}{\electronvolt}$ as in Ref. \onlinecite{Bistritzer11}. Details on the model are outlined in the Appendix \ref{App:CM}.

Experiments are usually done on a substrate of hexagonal boron-nitride (h-BN). Having a structure similar to graphene, an influence depending on the alignment with the substrate can be observed and  Kim et al. \cite{Kim2018} found that h-BN induces a band gap at the Dirac points. To account for this effect, we will introduce a general sublattice splitting with different bias parameter for the top ($\Delta_t$) and bottom ($\Delta_b$) layers as in Ref. \onlinecite{Bultinck19}.

In Fig. \ref{bandstructures:a}, the band structure around charge neutrality is shown for TBG in the presence of out-of-plane corrugation and sublattice splitting. The one-particle gap between the flat and remote bands is clearly seen at the $\Gamma$-point and also the substrate-induced splitting at the $K$-points can be appreciated.

\subsubsection{Tight-binding model}
Our study will be complemented by the same analysis based on the tight-binding model (TBM). Parameters are taken from Refs. \onlinecite{Brihuega12,Moon13} such that the nearest-neighbour intra-layer hopping parameter is set to $t=-2.7$ eV and the vertical interlayer hopping parameter to $t_\perp=0.48$ eV. 

As was the case in the CM, also for the TBM no clear single-particle gap appears that separates the flat from the remote bands. Thus, again lattice relaxation effects have to be taken into account and we choose the approach of Nam and Koshino.\cite{Nam17} To be more general, we will discuss two different parameterizations of the in-plane relaxation based on the original work\cite{Nam17} and updated parameters.\cite{Nam20} By this, we show that the different lattice relaxations only affect the analysis quantitatively, but not qualitatively. 

\begin{figure}[!t]
	\begin{subfigure}[b]{0.5\linewidth}
		\caption{ }\label{bandstructures:a}
		\hfill\includegraphics[width=0.95\textwidth]{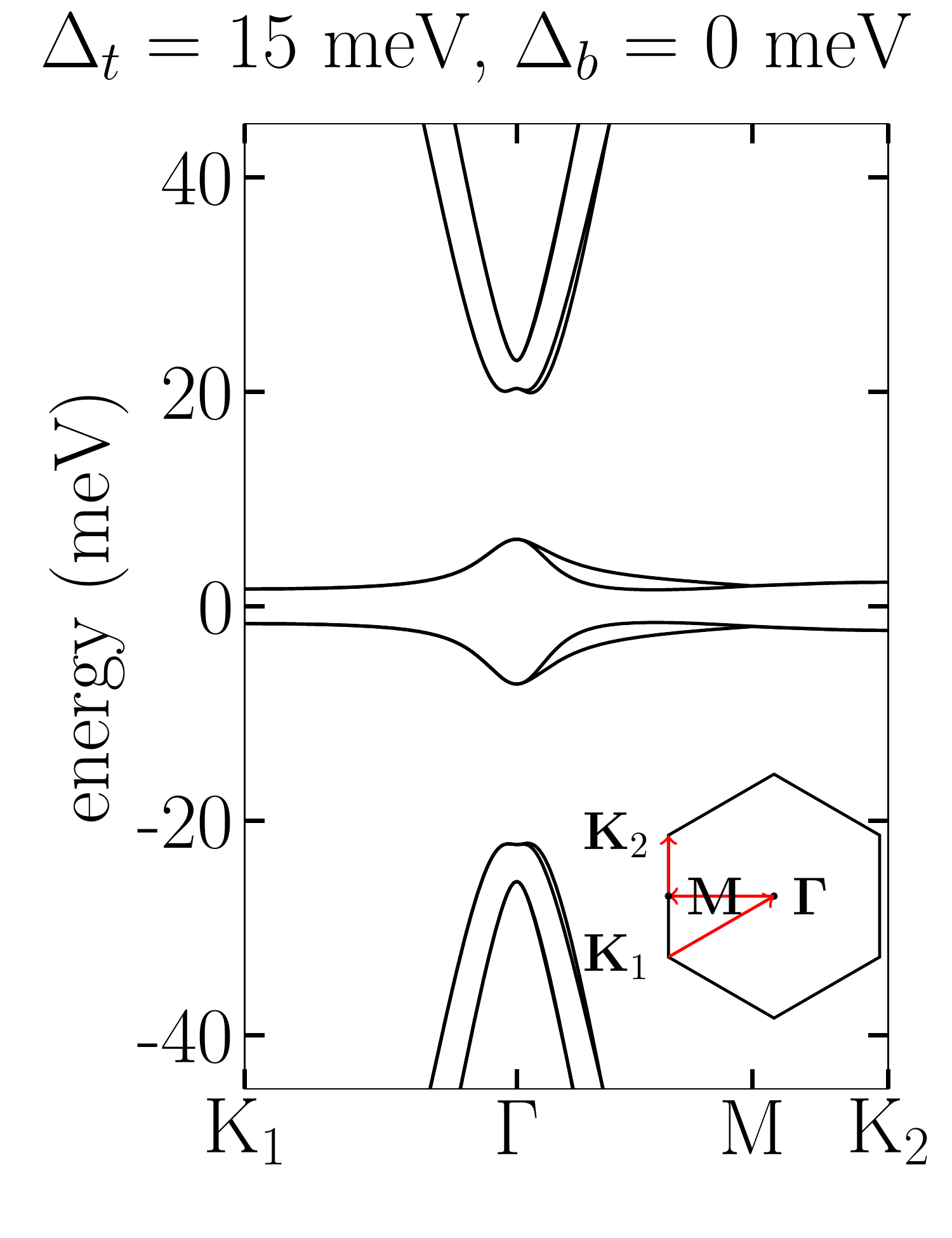}
	\end{subfigure}%
\hfill
	\begin{subfigure}[b]{0.5\linewidth}
		\caption{ }\label{bandstructures:b}
		\hfill\includegraphics[width=0.95\textwidth]{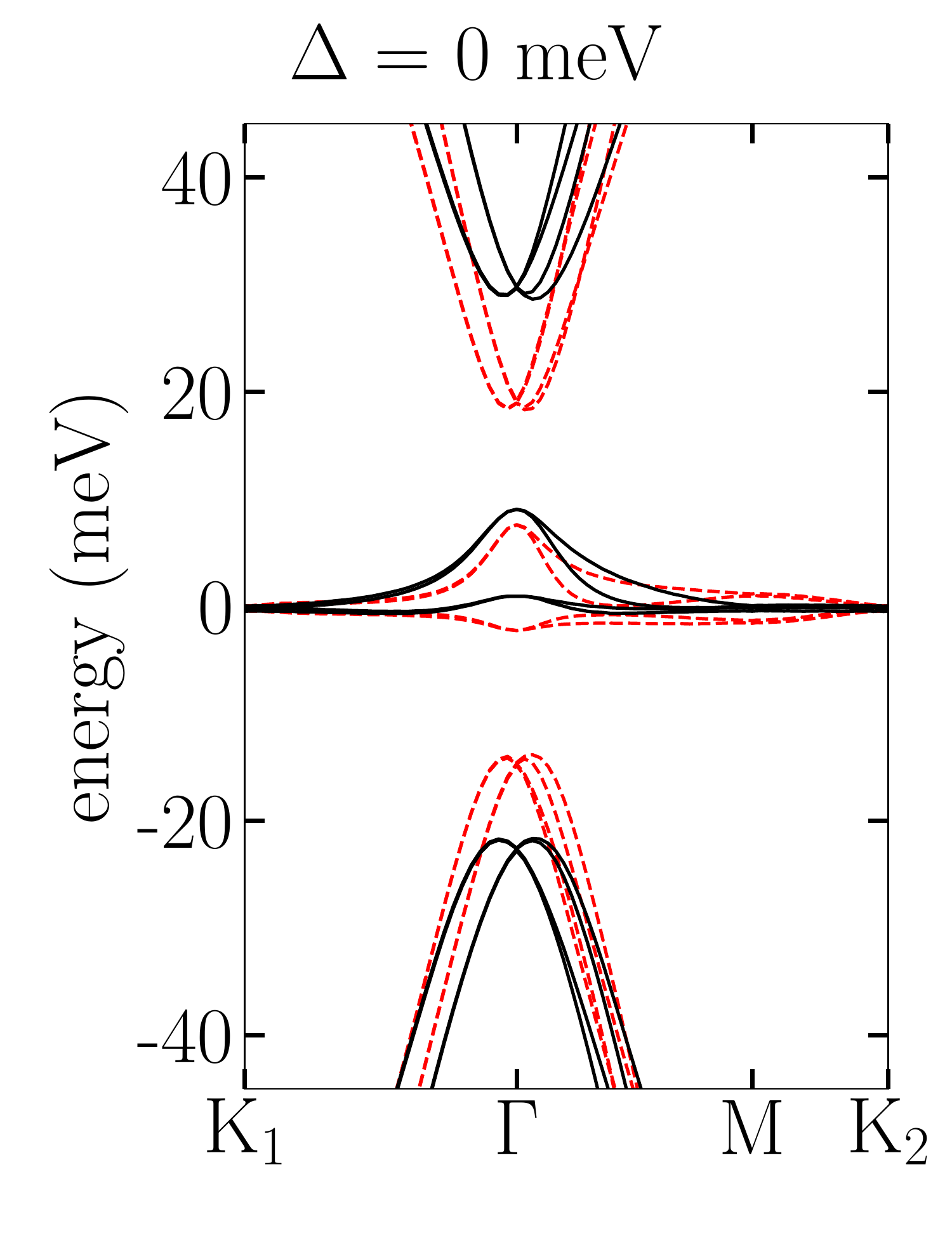}
	\end{subfigure}

	\caption{Bandstructures of TBG at the magic angle $\theta = 1.05^{\circ}$ around charge neutrality. (a) Continuum model (CM) with out-of-plane corrugation and a substrate induced sublattice splitting of $\Delta_t = \SI{15}{\milli\electronvolt}$ and $\Delta_b = \SI{0}{\milli\electronvolt}$. (b) Tight-binding model (TBM) with in-plane lattice relaxation for two parameter systems. The black curves stands to the updated relaxation parameters\cite{Nam20} and the red dashed curves are taken from Ref. \cite{Nam17} where the lattice relaxation was underestimated by a factor 0.42 relative to the actual lattice relaxation.}
	\label{fig:bandstructures}
\end{figure}

The resulting band structure can be seen in Fig. \ref{bandstructures:b} where the black curves refer to the updated ones to which we will from now on refer when talking of the relaxed TBM. The red dashed curves refer to the parameters of Ref. \cite{Nam17} where the lattice relaxation was underestimated by a factor 0.42 relative to the actual lattice relaxation. 

The influence of the substrate will also be discussed for the TBM and included in a way similar as in the CM. The on-site energy in the Hamiltonian in one layer is thus shifted to $\Delta$ for sites belonging to sublattice A and to $-\Delta$ for sublattice B. In contrary to the CM calculations, we will always neglect the sublattice-bias of the other layer zero.   

\subsection{Bandgap versus Bandwidth}
\begin{figure}[!t]
	\begin{subfigure}{0.85\linewidth}
		\caption{}\label{bandgap:a}
		\hfill\includegraphics[width=0.95\linewidth]{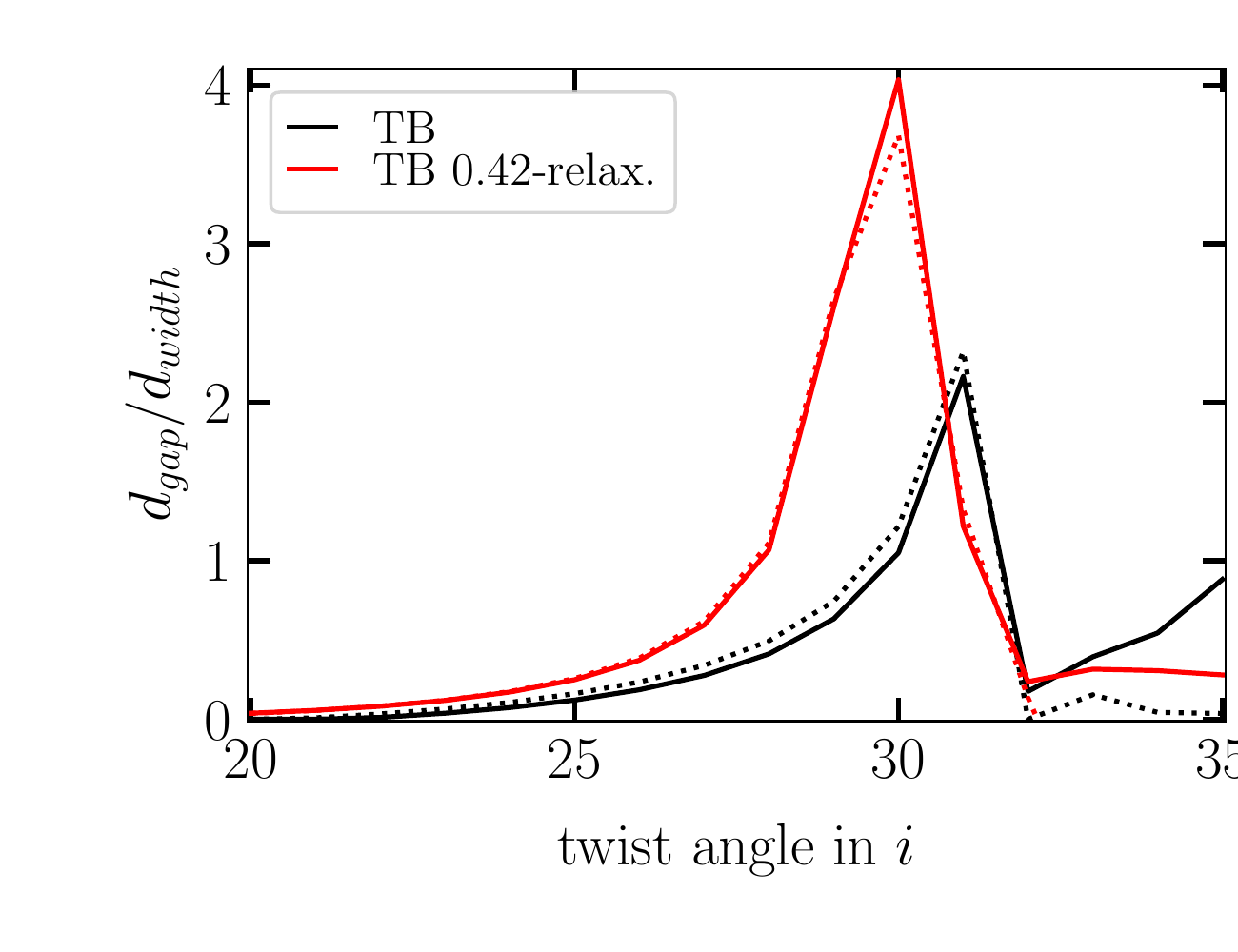}
	\end{subfigure}
	\begin{subfigure}{0.85\linewidth}
		\caption{}\label{bandgap:b}
		\hfill\includegraphics[width=0.95\linewidth]{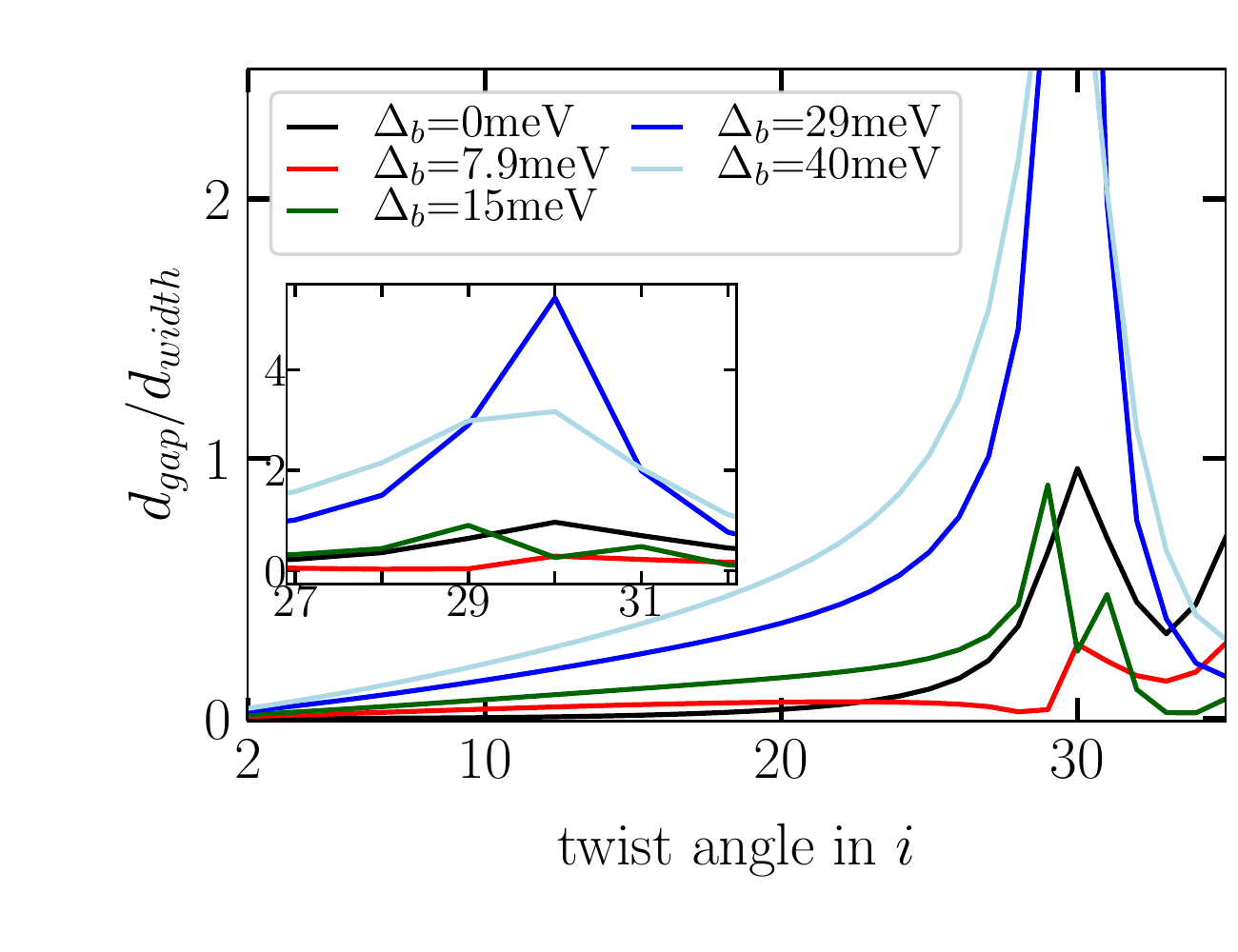}
	\end{subfigure}
	\caption{(a) Ratio of the gap between the flat and remote bands vs. the bandwidth of the flat bands. The solid (dashed) lines refer to the gap with respect to the upper (lower) remote bands.
		(b) Ratio of the gap per bandwidth, now referring to the gap between the lowest conduction and highest valence band compared to the respective conduction bandwidth. The gap at the Dirac cone was induced by the sublattice potential $\Delta_t = \SI{15}{\milli\electronvolt}$.}
	\label{fig:bandgap}
\end{figure}
Crucial for the application of the Mielke-Tasaki theorem is the flat-band condition which is only approximate in the case of TBG. In the following, we will, therefore, assess this condition quantitatively. 

The bands around charge neutrality can be regarded as nearly flat and separated from the rest of the spectrum, if the ratio of the bandgap to bandwidth $\frac{d_{gap}}{d_{width}}$ is large. This ratio is discussed in the following as function of the commensurate twist angle parametrized by the integer $i$. In this notation, the magic angle $\theta = 1.05^{\circ}$ corresponds to $i=31$. 

In Fig. \ref{bandgap:a}, the bandgap between the flat bands and the remote bands is shown as it turns out for the TBM. The sublattice bias $\Delta$ is set to zero in both cases. The curves of the TBM show a maximum around the magic angle supporting the claim that the bands flatten, decreasing the bandwidth and the relative width of the gap increases. 

In Fig. \ref{bandgap:b}, we discuss the ratio in the CM focusing on the additional gap that opens up in the presence of various sublattice biases. In all cases, the splitting opens up a gap between the two valence (lower) and two conduction (upper) flat bands at the Dirac point. Again, we observe a maximum around the magic angle. Interestingly, for fixed $\Delta_t=15meV$, there is an optical sublattice $\Delta_b\sim30$meV where the flat-band theorem can be applied.

\begin{figure}[!t]
	\begin{subfigure}{\linewidth}
		\caption{}\label{dm:a}
		\vspace*{11pt}
		\includegraphics[width=\linewidth]{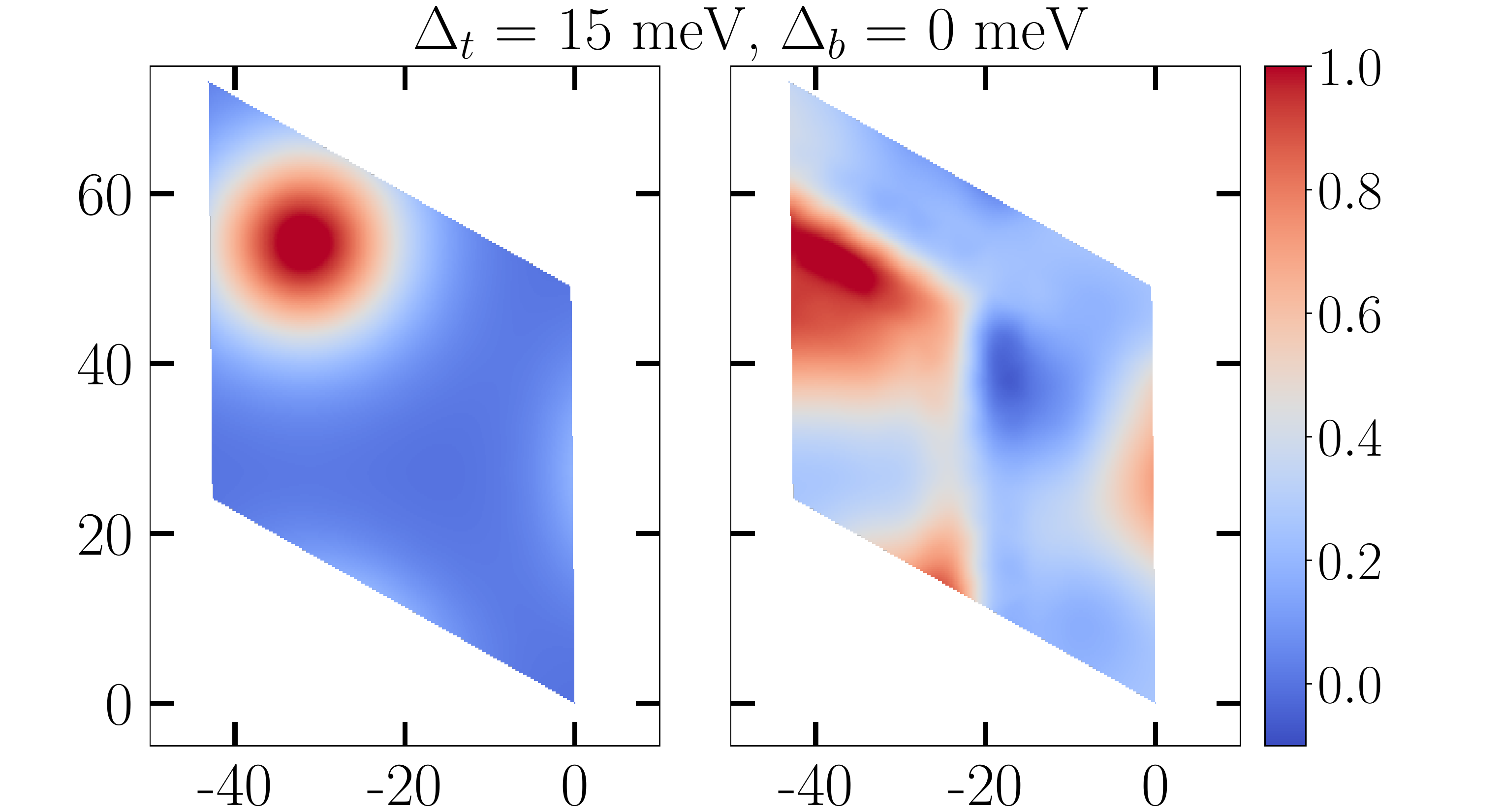}
	\end{subfigure}
	\begin{subfigure}{0.79\linewidth}
		\caption{}\label{dm:b}
		\hfill\includegraphics[width=0.95\linewidth]{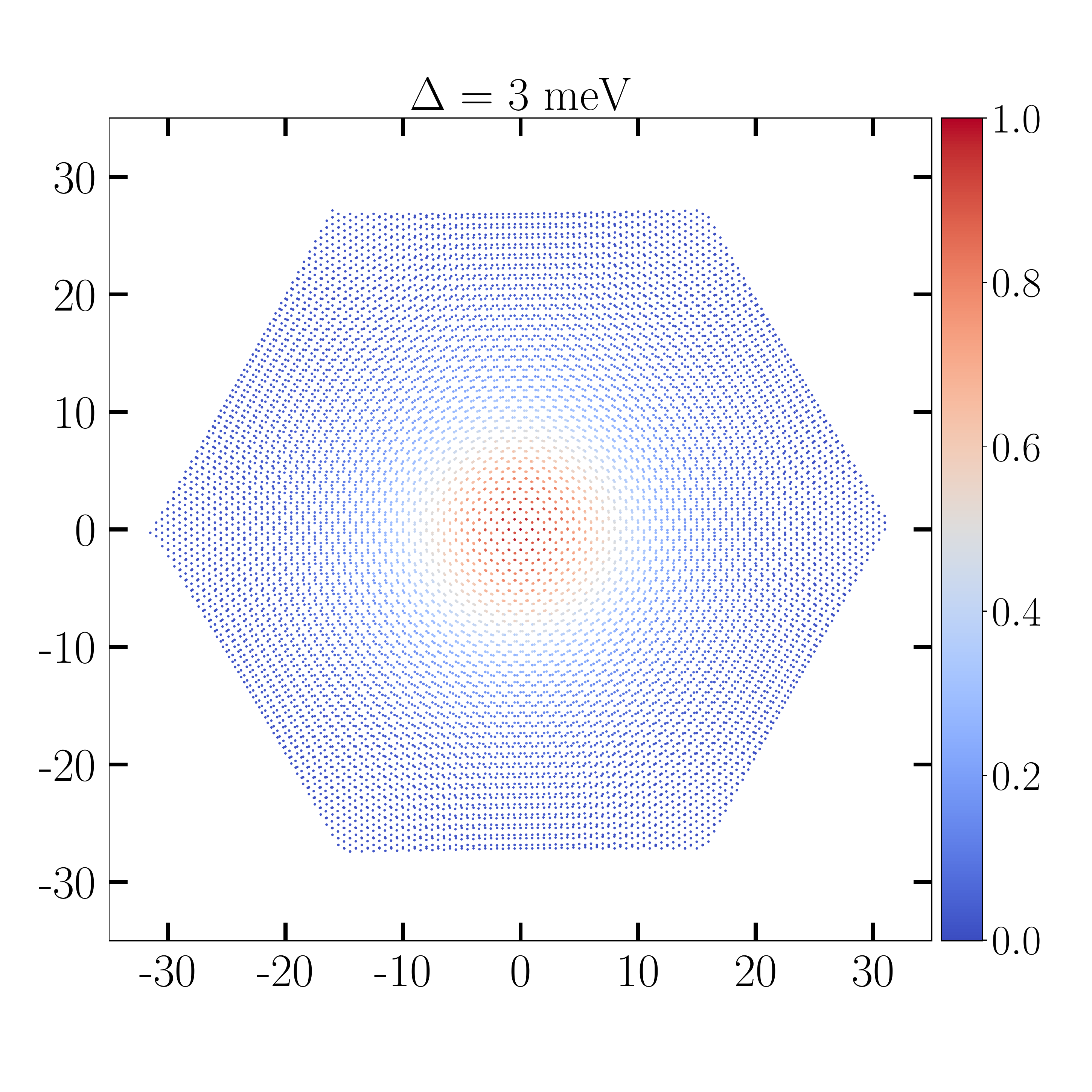}
	\end{subfigure}
	\begin{subfigure}{0.79\linewidth}
		\caption{}\label{dm:c}
		\hfill\includegraphics[width=0.95\linewidth]{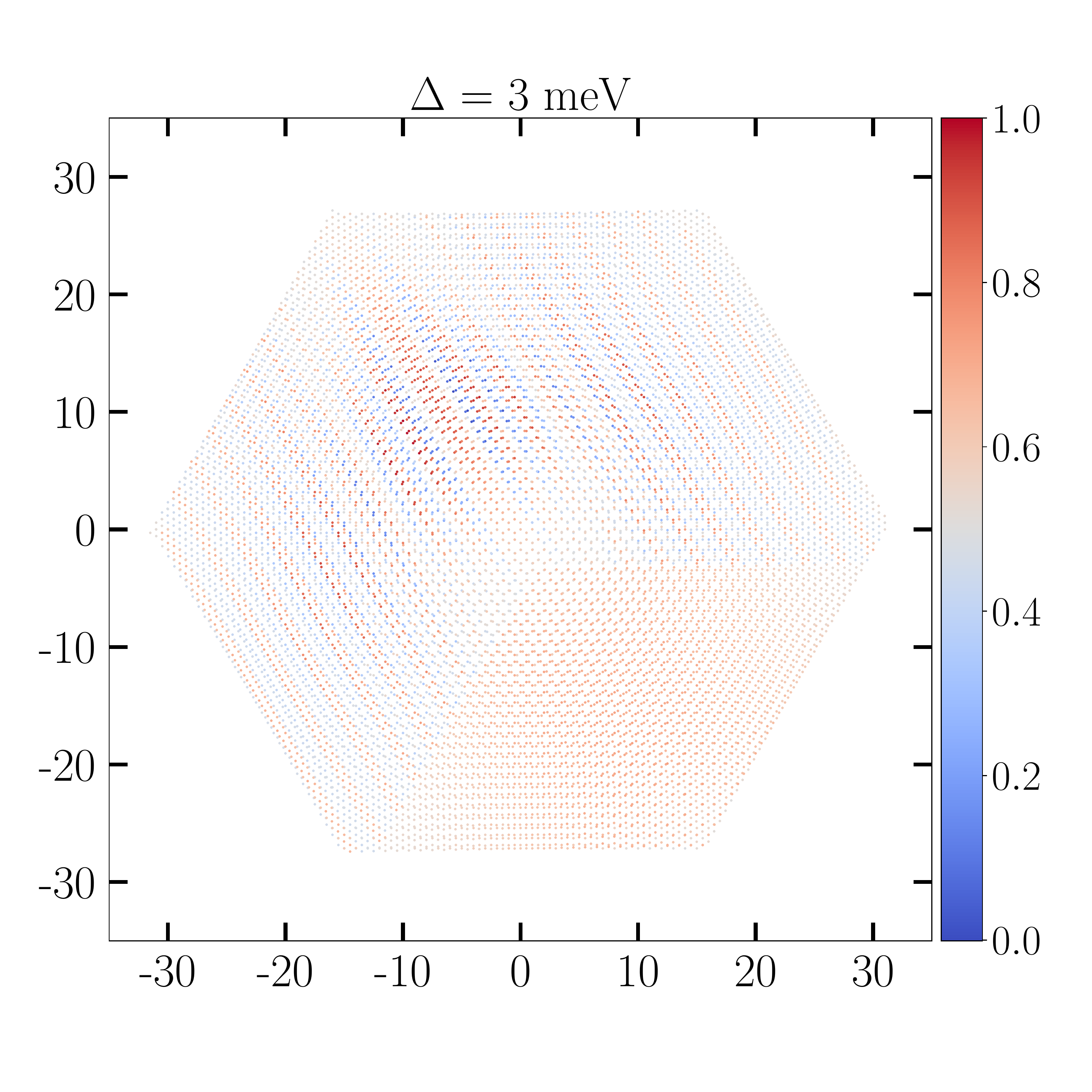}
	\end{subfigure}
	\caption{(a) Diagonal and off-diagonal component of the density matrix, $\rho_{ii}$ and  $\rho_{i(i+m)}$ with $m=35$, as obtained from the continuum model (CM) with the parameters of Fig. \ref{fig:bandstructures}. (b) Diagonal and (c) off-diagonal component with $m=35$ of the density matrix as obtained from the tight-binding calculations for a non-relaxed system with $\Delta =\SI{3}{\milli\electronvolt}$.}
	\label{fig:dm}
\end{figure}

\section{Density matrix of TBG}
We will now discuss the central quantity of our approach, the single-particle density matrix of the four and two bands around the neutrality point, respectively. 
\subsection{Density matrix of the CM}  
Within the CM, we solve the eigenvalue problem on an evenly spaced grid over the extended Brillouin zone, including both valleys. The density matrix $\rho_{ij}$ is then obtain from the following definition: 
\begin{align}
\label{eq:numeric-rho}
\rho_{ij} &= \frac{1}{A} \sum_n\sum_{\mathbf{k}\in 1.BZ} e^{i \, \mathbf{k}\cdot \left( \mathbf{R}_i - \mathbf{R}_j\right)}\\ 
&\times\sum_{\mathbf{G}_1 \mathbf{G}_2 } e^{i \, \left( \mathbf{G}_1 \cdot \mathbf{R}_i - \mathbf{G}_2 \cdot \mathbf{R}_j\right)} \boldsymbol{\Psi}^{\dagger}_{\mathbf{k+G}_1,n} \boldsymbol{\Psi}^{}_{\mathbf{k+G}_2,n}\notag
\end{align}
where the normalization constant is given by $A = N_{\mathbf{k}} N_{\mathbf{G}}^2$ and $N_{\mathbf{k}}$ denotes the number of $\mathbf{k}$-points, whereas $N_{\mathbf{G}}$ is the number of reciprocal vectors included in the calculation. $\mathbf{R}_i$ and $\mathbf{R}_j$ are the real space lattice points considered and the $n$th-eigenstate at $\mathbf{k+G}$ is denoted by $\boldsymbol{\Psi}_{\mathbf{k+G},n}$. The sum over $n$ runs either over the 4 states around the neutrality point or the 2 highest valence/2 lowest conduction band states, respectively. 

The density matrix is defined on a coarsed grained unit cell and usually $\sim100$ points are sufficient to resolve the main features. In Fig. \ref{dm:a}, we display the diagonal elements $\rho_{ii}$ on the rhombic unit cell as well as an off-diagonal element $\rho_{i(i+m)}$ with $m=35$. The diagonal elements are characterized by a clear maximum at the AA-stacked region which is smeared out in the off-diagonal elements.

\subsection{Density matrix of the tight-binding model} 
We also calculate the density matrix with respect to the tight-binding model. In this case, $\rho_{ij}$ is given by the following formula:
\begin{equation}
\label{eq:numeric-rho-TB}
\rho_{ij} = \frac{1}{N_{\mathbf{k}}} \sum_n\sum_{{\mathbf{k}}\in1.BZ}  \boldsymbol{\Psi}^{\dagger}_{\mathbf{k},n}(\mathbf{R}_i)\boldsymbol{\Psi}_{\mathbf{k},n}(\mathbf{R}_j)\;,
\end{equation}
where $N_{\mathbf{k}}$ denotes the number of $\mathbf{k}$-points included in the calculation and $\boldsymbol{\Psi}_{\mathbf{k},n}(\mathbf{R}_i)$ is the component at $\mathbf{R}_i$ of the $n$th-eigenstate at $\mathbf{k}$.

In Fig. \ref{dm:b}, the diagonal elements of the density matrix for a non-relaxed system with $\Delta =\SI{3}{\milli\electronvolt}$ is plotted. In the case of the tight-binding model, the unit cell shall be resolved by the atomistic lattice sites, i.e., for $i=31$, the unit cell contains 11908 atoms and the density matrix thus has dimensions of $11908\times11908$. Also for this model, the diagonal components show a clear maximum at the AA-stacked regions. In Fig. \ref{dm:c}, we show the off-diagonal element $\rho_{i(i+m)}$ for $m=35$. Again, the diagonal matrix elements are characterized by a clear maximum at the AA-stacked region which is smeared out in the off-diagonal elements.

\section{Irreducibility} 
For both models, we have calculated the density matrix according to Eqs. (\ref{eq:numeric-rho}) and (\ref{eq:numeric-rho-TB}), respectively. In order to show flat-band ferromagnetism following Mielke's theorem, we need to show that  those matrices are irreducible. An irreducible matrix is often defined by the matrix not being reducible. Since we deal only with Hermitian matrices, a sufficient condition for a matrix to be reducible is that there exists a permutation of columns and rows that transform the matrix into a block-diagonal form $\left( \begin{smallmatrix} \mathbf{A}_{11} & \mathbf{0} \\ \mathbf{0} & \mathbf{A}_{22} \end{smallmatrix} \right)$. Instead of proving the non-existence of such a permutation 
we employ the equivalent, but more direct, definition from graph theory: A matrix $\rho \in \mathbb{R}^{(n,n)}$, $\rho \neq 0$ is irreducible if and only if the corresponding adjacency graph is connected which shall be discussed below.

The numerical implementation of the test for irreducibility consists of two steps. First, $\rho$ is transformed into its adjacency matrix $\widehat{\rho}$, then the code tests whether $\widehat{\rho}$'s graph is connected. For our purposes, the transformation from $\rho$ into $\widehat{\rho}$ slightly differs from the usual textbook (e.g. Ref.\onlinecite{Kna2018})definition (where it would read "$1 , \, \text{ if } \rho_{ij}\neq 0$"):
\begin{equation}
\label{eq:adjacency}
\widehat{\rho}_{ij} \coloneqq 
\begin{cases}
1 , & \text{ if } \rho_{ij} \geq \tau \\
0 , & \text{otherwise}
\end{cases}
\end{equation}
where the threshold $\tau$ is a variable and can be set. By choosing $\tau$ finite, we can probe how stable the graph is connected, but we can also compensate for numerical errors that do not allow to simply set $\tau=0$. If, with a threshold higher than the numerical error, $\widehat{\rho}$ is still irreducible, one can assume that $\rho_{ij}$ itself is irreducible. In the following, we will determined a "critical" threshold $\tau_c$ which we define as the largest $\tau$ that can be set before $\widehat{\rho}$ becomes reducible. 

The second step in proving the graph's connectedness is done by a path finding algorithm. The graph is connected if from every node $v_i$ $i \in \{ 1, \ldots , N \}$ every other node can be reached. Going to the assigned adjacency matrix conserves this symmetry. Density matrices are symmetric and  thus is $\widehat{\rho}$. In the graph of a symmetric matrix, every connection would exist in both directions. Therefore, the algorithm only needs to find paths between nodes in one direction $v_i \rightarrow v_j$ and it immediately follows $v_j \rightarrow v_i$. Also, from $v_j \rightarrow v_i$ follows $v_j \rightarrow v_k$ $\forall k \in \{ 1, \ldots , N \}$ if there exists a path $v_i \rightarrow v_k$. Thus, the graph being connected is equivalent to $\exists (v_1 \rightarrow v_k)$ $\forall k \in \{ 1, \ldots , N \}$. 

For the  code implementation, the problem is further reduced to the question whether there exists an edge $(v_1, v_k)$ or alternatively an edge $(v_k,v_j)$ with $v_1 \rightarrow v_j$. This can be treated recursively. The algorithm first finds all nodes $v_i$, $i \in I \subset \{ 2, \ldots , N \}$  with edges $(v_1,v_i)$ and adds them to the set of reachable nodes $M_{found} = \{ 1\} \oplus I$. In the next step, all nodes $v_j$ are found that have an edge $(v_i,v_j)$ and $j \neq 1 \wedge j \notin I$. Those $j$ are included in $M_{found}$, $M_{found} = \{ 1\} \oplus I \oplus J$. The following step starts with $v_j$ and the scheme is repeated until $M_{found} = \{ 1, \ldots , N \}$ or until the element last added to $M_{found}$ has no edge with any node not yet in $M_{found}$. The way the algorithm works in the latter case implies $\not\exists (v_i,v_j)$ with $i \in M_{found}$ and $j \notin M_{found}$. Thus, the graph is not connected and $\rho_{ij}$ reducible. $M_{found} = \{ 1, \ldots , N \}$  means that all nodes have been reached, so the graph is fully connected and $\rho_{ij}$ irreducible.

\begin{figure}[!t]
	\includegraphics[width=1.05\linewidth]{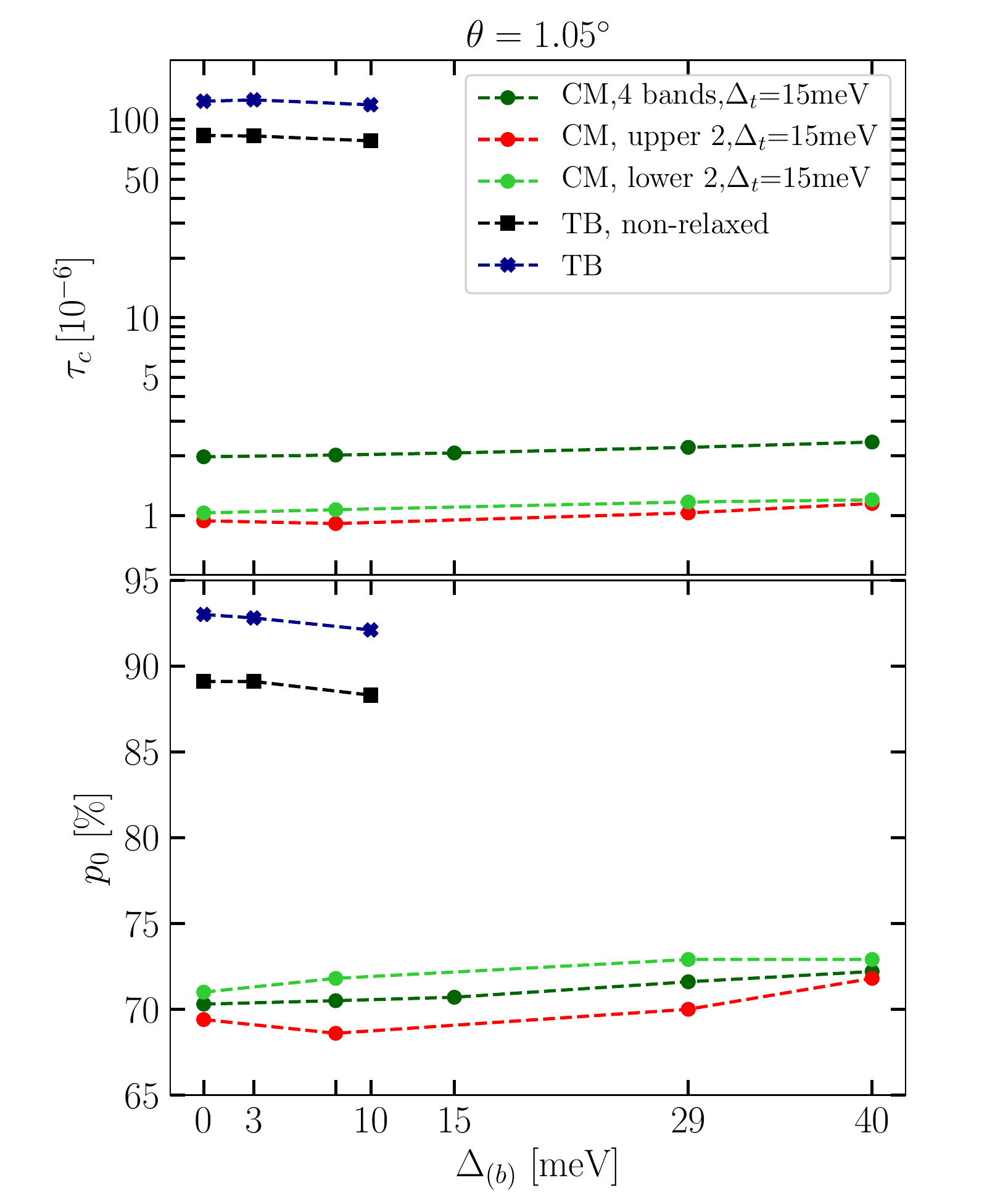}
	\caption{Thresholds $\tau_c$ for the density matrix at the magic angle $\theta = 1.05^{\circ}$ with different models as function of the sublattice splitting $\Delta$ (TBM) or of $-\Delta_b$ (CM). The upper plot displays the actual value $\tau_c$ while the lower plot shows the percentage $p_0$ of the matrix elements that have been set to zero by this threshold.}
	\label{fig:thresholds1}
\end{figure}
\section{Results} 
Having detailed our methods, we first present our results in form of Table 1 and 2 for the CM and the TBM, respectively. For comparison, we also analyzed the non-relaxed lattices that do not show a gap between the flat and remote bands. The column "bands" details whether the calculation runs on all 4 flat bands or only on the conduction (upper) or valence (lower) bands. This always implies that the included flat bands are half filled: "4" thus means ferromagnetism at charge neutrality, whereas "lower 2" or "upper 2" implies ferromagnetism at half-filling of the valence and conduction band, respectively. 

We also considered the unrelaxed lattice and a sublattice gap $\Delta$ when we are only interested in the ground state at the neutrality point. This shows that the irreducibility of the matrix does not depend on the particular choice of parameters. 

\begin{table}[h]
		\begin{minipage}{\linewidth}
		\resizebox{\linewidth}{!}{%
		\begin{tabular}{l|c|c||c||c|c|r}
			relaxed & $\Delta_t$ $\left[ meV \right]$ & $\Delta_b$ $\left[ meV \right]$  & bands & $\tau_c$ & $p_0$ $ \left[ \% \right]$ \\ \hline
			no & 0 & 0 & 4 & $1.22 \times 10^{-6}$ & 72.1 \\ \hline
			\multirow{1}{*}{yes} & \multirow{1}{*}{0} & \multirow{1}{*}{0} & 4 & $1.94 \times 10^{-6}$ & 70.0 \\\hline
			\multirow{3}{*}{yes} & \multirow{3}{*}{15} & \multirow{3}{*}{0} & 4 & $1.98 \times 10^{-6}$ & 70.3 \\
			& & & upper 2 & $9.4  \times 10^{-7}$ & 69.4 \\
			& & & lower 2 & $1.03 \times 10^{-6}$ & 71.0 \\ \hline
			\multirow{3}{*}{yes} & \multirow{3}{*}{15} & \multirow{3}{*}{-7.9} & 4 & $2.02 \times 10^{-6}$ & 70.5 \\
			& & & upper 2 & $9.1  \times 10^{-7}$ & 68.6 \\
			& & & lower 2 & $1.07 \times 10^{-6}$ & 71.8 \\ \hline
			\multirow{1}{*}{yes} & \multirow{1}{*}{15} & \multirow{1}{*}{-15} & 4 & $2.07 \times 10^{-6}$ & 70.7 \\ \hline
			\multirow{3}{*}{yes} & \multirow{3}{*}{15} & \multirow{3}{*}{-29} & 4 & $2.21 \times 10^{-6}$ & 71.6 \\
			& & & upper 2 & $1.03  \times 10^{-6}$ & 70.0 \\
			& & & lower 2 & $1.17 \times 10^{-6}$ & 72.9 \\ \hline
			\multirow{3}{*}{yes} & \multirow{3}{*}{15} & \multirow{3}{*}{-40} & 4 & $2.35 \times 10^{-6}$ & 72.2 \\
			& & & upper 2 & $1.15  \times 10^{-6}$ & 71.8 \\
			& & & lower 2 & $1.20 \times 10^{-6}$ & 72.9 
	\end{tabular}}

\end{minipage}	
	\label{tab:result1}
	\caption{Irreducibility analysis based on the density matrix obtained from the continuum model at the magic angle $\theta = 1.05^{\circ}$ ($i=31$). For all parameters, the critical values are well above the numerical accuracy and the corresponding density matrix is thus irreducible.}
\end{table}

\begin{table}[h]
	\begin{tabular}{l|c||c|c|r}
		relaxed & $\Delta$ $\left[ meV \right]$ & bands & $\tau_c$ & $p_0$ $ \left[ \% \right]$ \\ \hline
		no & 0 & 4 &$8.32 \times 10^{-5}$ & 89.1 \\
		yes & 0 & 4 & $1.24 \times 10^{-4}$ & 93.0 \\
		no & 3 & 4 &$8.28 \times 10^{-5}$ & 89.1 \\
		yes & 3 & 4 &$1.26 \times 10^{-4}$ & 92.8 \\
		no & 10 & 4 &$7.82 \times 10^{-5}$ & 88.3 \\
		yes & 10 & 4 &$1.19 \times 10^{-4}$ & 92.1 
		
	\end{tabular}
	\label{tab:result2}
	\caption{Irreducibility analysis based on the density matrix obtained from the tight-binding model at the magic angle $\theta = 1.05^{\circ}$ ($i=31$). For all parameters, the critical values are well above the numerical accuracy and the corresponding density matrix is thus irreducible.}
\end{table}

\begin{figure}[h]
	\includegraphics[width=1.05\linewidth]{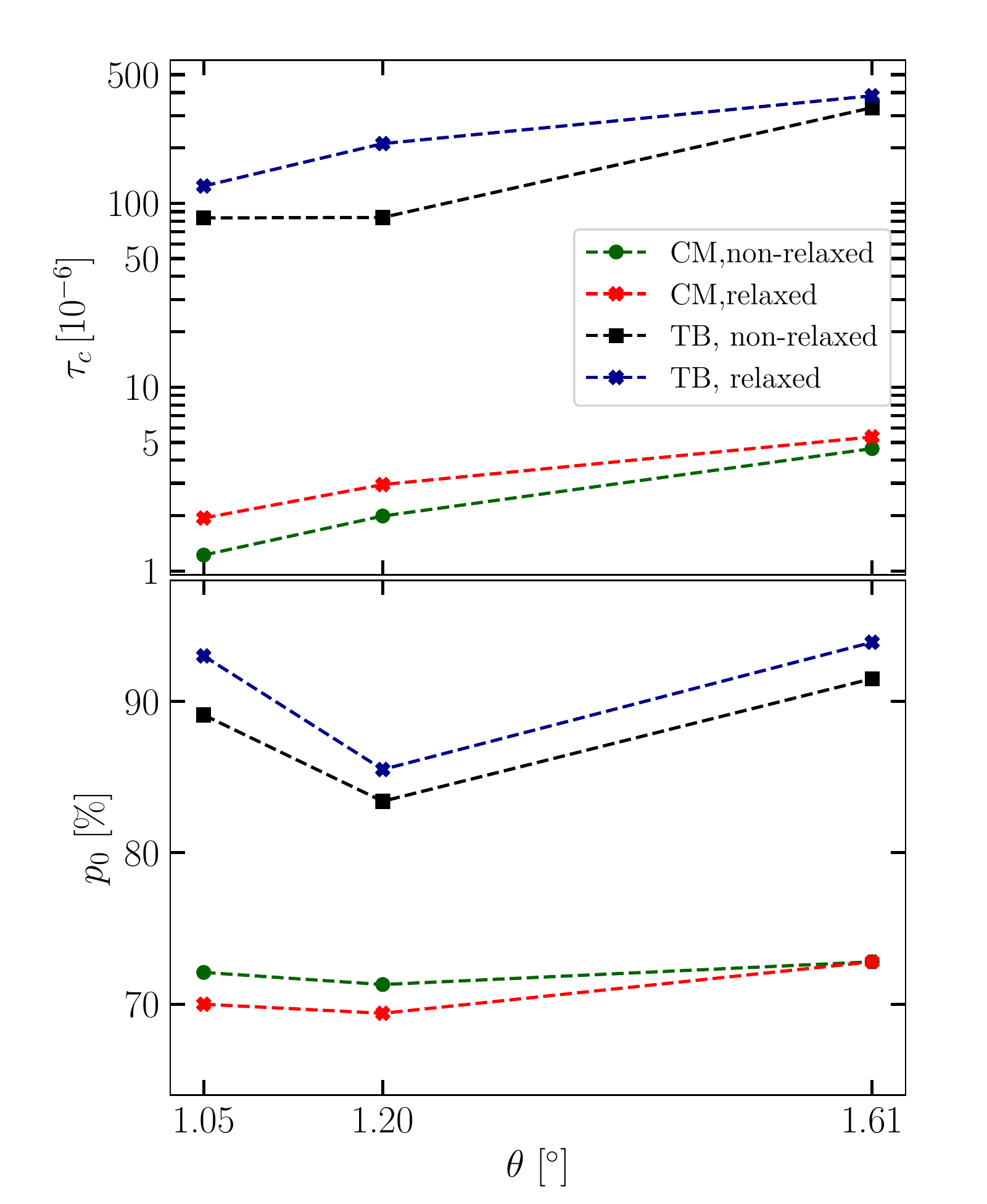}
	\caption{Thresholds $\tau_c$ for different angles in CM and commensurate tight-binding model without sublattice splitting.}
	\label{fig:thresholds2}
\end{figure}

In Figs. \ref{fig:thresholds1} and \ref{fig:thresholds2}, we present a summary of our results graphically. Additional results for the irreducibility also for other angles can be found in the Appendix \ref{App:Irreducibility}. They confirm our main conclusion that the critical values $\tau_c$ are much larger than the expected numerical errors. This holds, first of all, for the case where all four bands are considered and we expect ferromagnetism at half-filling. But it also holds for the case where only two bands were considered, referring to ferromagnetism at half-filling.  

\section{Discussion and Outlook}
We investigated the single particle density matrix of the almost flat bands of TBG around charge neutrality. Our main conclusion is that  the single particle density matrix is irreducible for virtually all parameters. This is the main condition for flat band ferromagnetism to appear. \cite{Mielke93,Mielke1999} Clearly, this does not prove the appearance of ferromagnetism in TBG in a mathematical sense. But we argue that nevertheless one should expect flat band ferromagnetism for the following reasons:

(i) The bands in TBG are not completely flat but there is a sufficiently large gap in the spectrum. For many other system it has been shown that flat band ferromagnetism is robust against a small dispersion of the flat band in that case and if the interaction is not too small.

(ii) In the mathematical proofs of flat band ferromagnetism the flat band needs to appear either on the bottom or on the top of the spectrum or the lattice needs to be bipartite. But one can use a perturbational argument and a conjecture on the monotonicity of $S^2$  as a function of $U$ to argue that flat band ferromagnetism is not restricted to these cases but can be expected in a much wider range of models and systems including TBG.

(iii) There may be further interactions present in TBG but one might expect that the Hubbard model describes the essential physics of TBG. Further, additional interactions do not necessarily disturb flat band ferromagnetism.\cite{Strack93a,Kollar01}\\

{{\it Note after proof:} During completion of the manuscript, we became aware of Ref. \cite{Wu20} that contains a similar conclusion as ours.} \\

\section{Acknowledgments} This work has been supported by Spain's MINECO under Grant No. FIS2017-82260-P, by Germany's Deutsche Forschungsgemeinschaft (DFG) via SFB 1277 as well as by the CSIC Research Platform on Quantum Technologies PTI-001.
\appendix
\section{Hamiltonian of the continuum model} 
\label{App:CM}
The CM's full Hamiltonian is given by the Hamiltonian of the single unrotated layer, $H^{\mathbf{K}}_D$, the one of the single rotated layer, $H^{\mathbf{K}^{\theta}}_D$, and the interlayer coupling, $H_T$:
\begin{equation}
H_{CM}=H^{\mathbf{K}}_D+H^{\mathbf{K}^{\theta}}_D+H_T
\end{equation}
This yields the non-zero  matrix elements
\begin{align}
&\braket{\Psi^{(1)}_{(\mathbf{K+q}), \alpha} | H_{CM} | \Psi^{(2)}_{(\mathbf{K}^{\theta}+\mathbf{s}^{\theta}), \beta}}=T^{\alpha \beta}_{\mathbf{q}\mathbf{s}^{\theta}} \\\notag
&=  T^{\alpha \beta}_b \delta_{\mathbf{q-s}^{\theta} , \mathbf{q}_b} + T^{\alpha \beta}_{tr} \delta_{\mathbf{q-s}^{\theta} ,\mathbf{q}_{tr} } + T^{\alpha \beta}_{tl} \delta_{\mathbf{q-s}^{\theta} , \mathbf{q}_{tl}}
\end{align}
the corresponding hermitian conjugates and
\begin{equation}
\begin{split}
&\braket{\Psi^{(1)}_{(\mathbf{K+q}), \alpha} | H_{CM} | \Psi^{(1)}_{(\mathbf{K+s}), \beta}} \\
&= \delta_{\mathbf{q} , \mathbf{s}} \left( H^{\mathbf{K}}_D(\mathbf{q})\right)^{\alpha \beta} \;,
\end{split}
\end{equation}
\begin{equation}
\begin{split}
&\braket{ \Psi^{(2)}_{(\mathbf{K}^{\theta}+\mathbf{q}^{\theta}), \alpha} | H_{CM} | \Psi^{(2)}_{(\mathbf{K}^{\theta}+\mathbf{s}^{\theta}), \beta} }\\ &= \delta_{\mathbf{q}^{\theta} , \mathbf{s}^{\theta}} \left( H^{\mathbf{K}^{\theta}}_D(\mathbf{q}^{\theta})\right)^{\alpha \beta}\;,
\end{split}
\end{equation}
where $\left( H^{\mathbf{K}}_D(\mathbf{q})\right)^{\alpha \beta}$ and $\left( H^{\mathbf{K}^{\theta}}_D(\mathbf{q}^{\theta})\right)^{\alpha \beta} $ are elements of the matrices
\begin{equation}
\label{eq:H-Dirac1}
H^{\pm \mathbf{K}}_D(\mathbf{q}) = \pm v_F \left| \mathbf{q} \right|
\begin{pmatrix}
0 & e^{\mp i \theta_{\mathbf{q}}} \\
e^{\pm i \theta_{\mathbf{q}}} & 0
\end{pmatrix}\;,
\end{equation}
\begin{equation}
\label{eq:H-Dirac2}
H^{\pm \mathbf{K}^{\theta}}_D(\mathbf{q}) = \pm v_F \left| \mathbf{q} \right|
\begin{pmatrix}
0 & e^{\mp i (\theta_{\mathbf{q}} - \theta)} \\
e^{\pm i (\theta_{\mathbf{q}} - \theta)} & 0
\end{pmatrix}\;.
\end{equation}
The different signs stand for the different valleys and $T_{b}$, $T_{tr}$, and $T_{tl}$ are\cite{Koshino18}
\begin{equation}
\begin{split}
T_{b} = 
\begin{pmatrix}
u & u^{\prime} \\
u^{\prime} & u
\end{pmatrix} , \qquad
T_{tr} =  
\begin{pmatrix}
u e^{i \phi} & u^{\prime} \\
u^{\prime} e^{-i \phi} & u e^{i \phi}
\end{pmatrix} , \\
T_{tl} = \omega 
\begin{pmatrix}
u e^{-i \phi} & u^{\prime} \\
u^{\prime} e^{i \phi} & u e^{-i \phi}
\end{pmatrix}
\end{split}
\end{equation}
with $\phi = \frac{2 \pi}{3}$.\\ \\

We will also introduce a general sublattice splitting with one bias parameter for the top layer $\Delta_t$ and one bias for the bottom layer $\Delta_b$.\cite{Bultnick19} The interlayer part of the Hamiltonian is expanded by  $\Delta_t \sigma^z$ and $\Delta_b \sigma^z$, respectively, reading now
\begin{equation}
\braket{\Psi^{(1)}_{(\mathbf{K+q}), \alpha} | H_{CM} | \Psi^{(1)}_{(\mathbf{K+s}), \beta}}  
= \delta_{\mathbf{q} , \mathbf{s}}  v_F \left| \mathbf{q} \right|
\begin{pmatrix}
\Delta_b & e^{- i \theta_{\mathbf{q}}} \\
e^{ i \theta_{\mathbf{q}}} & -\Delta_b
\end{pmatrix}
\end{equation}
\begin{align}
\begin{split}
&\braket{ \Psi^{(2)}_{(\mathbf{K}^{\theta}+\mathbf{q}^{\theta}), \alpha} | H_{CM} | \Psi^{(2)}_{(\mathbf{K}^{\theta}+\mathbf{s}^{\theta}), \beta} } \\&= \delta_{\mathbf{q}^{\theta} , \mathbf{s}^{\theta}} v_F \left| \mathbf{q} \right|
\begin{pmatrix}
\Delta_t & e^{- i (\theta_{\mathbf{q}} - \theta)} \\
e^{ i (\theta_{\mathbf{q}} - \theta)} & -\Delta_t
\end{pmatrix}
\end{split}
\end{align}
\section{More results on the irreducibility of the single particle density matrix} 
\label{App:Irreducibility}
In this Appendix, we will give more details on our irreducibility analysis. In these more extensive tables, we also list numerical parameters such as the grid size $N_{\mathbf{k}}$ of the 1. Brillouin zone and the reciprocal lattice truncation $N_{\mathbf{G}}$, i.e., the number of included reciprocal lattice vectors. 

For the continuum model, we also include $N_{\mathbf{R}}$ which is the number of real space points $\mathbf{R}$ used to represent the density matrix $\rho_{ij}$. $N_{bands}$ is the number of bands included where (i) "2" means two valence bands {\it or} two conduction bands, and (ii) "4" means two valence {\it and} two conduction bands. In both cases, the valley degree of freedom is included, whereas the spin degree of freedom is ignored. 

Let us first present our results from the CM. Table \ref{tab:res1} contains our analysis for the non-relaxed and Table \ref{tab:res2} for the relaxed lattice of TBG, also including different sublattice biases. In both cases, 4 bands are considered predicting a ferromagnetic ground state at charge neutrality. In Table \ref{tab:res3} and \ref{tab:res3}, we analyze the system for  half-filled flat valence and conduction bands, respectively. 

\begin{table}[h]
	\begin{minipage}{\linewidth}
	\resizebox{\linewidth}{!}{%
	\begin{tabular}{l||c|c|c|c||c|r}
		$\theta$ $\left[ ^{\circ}\right]$ & $N_{\mathbf{G}}$ & $N_{\mathbf{k}}$ & $N_{\mathbf{R}}$ & $N_{bands}$ & $\tau$ & $p_0$ $ \left[ \% \right]$ \\ \hline
		0.93 & 9 & 324 & 100 & 4 & $1.29 \times 10^{-6}$ & 77.5 \\
		1.05 & 9 & 8100 & 100 & 4 & $1.09 \times 10^{-6}$ & 71.7 \\
		1.05 & 9 & 324 & 100 & 4 & $1.22 \times 10^{-6}$ & 72.1 \\
		1.05 & 9 & 81 & 100 & 4 & $1.47 \times 10^{-6}$ & 73.2 \\
		1.12 & 9 & 324 & 100 & 4 & $1.47 \times 10^{-6}$ & 71.6 \\
		1.12 & 9 & 324 & 400 & 4 & $1.47 \times 10^{-6}$ & 69.3 \\
		1.20 & 9 & 324 & 100 & 4 & $1.99 \times 10^{-6}$ & 71.3 \\
		1.61 & 9 & 324 & 100 & 4 & $4.63 \times 10^{-6}$ & 72.8
	\end{tabular}}
\end{minipage}	
	\caption[Thresholds non-relaxed systems]{Threshold values for the non-relaxed TBG modeled by the CM with $u^{\prime}=u=\SI{0.11}{\electronvolt}$.}\label{tab:res1}
\end{table}

\begin{table}[h]
	\begin{minipage}{\linewidth}
	\resizebox{\linewidth}{!}{%
	\begin{tabular}{l|c|c||c|c|c|c||c|r}
		$\theta$ $\left[ ^{\circ}\right]$ & $\Delta_t$ $\left[ meV \right]$ & $\Delta_b$ $\left[ meV \right]$  & $N_{\mathbf{G}}$ & $N_{\mathbf{k}}$ & $N_{\mathbf{R}}$ & $N_{bands}$ & $\tau$ & $p_0$ $ \left[ \% \right]$ \\ \hline
		1.61 & 0 & 0 & 9 & 324 & 100 & 4 & $5.36 \times 10^{-6}$ & 72.8 \\
		1.20 & 0 & 0 & 9 & 324 & 100 & 4 & $2.95 \times 10^{-6}$ & 69.4 \\
		1.05 & 0 & 0 & 9 & 324 & 100 & 4 & $1.94 \times 10^{-6}$ & 70.0 \\
		1.05 & 15 & 0 & 9 & 324 & 100 & 4 & $1.98 \times 10^{-6}$ & 70.3 \\
		1.05 & 15 & -7.9 & 9 & 324 & 100 & 4 & $2.02 \times 10^{-6}$ & 70.5 \\
		1.05 & 15 & -15 & 9 & 324 & 100 & 4 & $2.07 \times 10^{-6}$ & 70.7 \\
		1.05 & 15 & -29 & 9 & 324 & 100 & 4 & $2.21 \times 10^{-6}$ & 71.6 \\
		1.05 & 15 & -40 & 9 & 324 & 100 & 4 & $2.35 \times 10^{-6}$ & 72.2 
	\end{tabular}}	
\end{minipage}
	\caption[Thresholds relaxed systems]{Threshold values for the relaxed lattice with $u = \SI{0.0898}{\electronvolt}$, $u^{\prime} = \SI{0.11}{\electronvolt}$ modeled by the CM.}\label{tab:res2}
\end{table}

\begin{table}[h]
	\begin{minipage}{\linewidth}
	\resizebox{\linewidth}{!}{%
	\begin{tabular}{l|c|c||c|c|c|c||c|r}
		$\theta$ $\left[ ^{\circ}\right]$ & $\Delta_t$ $\left[ meV \right]$ & $\Delta_b$ $\left[ meV \right]$  & $N_{\mathbf{G}}$ & $N_{\mathbf{k}}$ & $N_{\mathbf{R}}$ & $N_{bands}$ & $\tau$ & $p_0$ $ \left[ \% \right]$ \\ \hline
		1.05 & 15 & 0 & 9 & 324 & 100 & 2 & $1.03 \times 10^{-6}$ & 71.0 \\
		1.05 & 15 & -7.9 & 9 & 324 & 100 & 2 & $1.07 \times 10^{-6}$ & 71.8 \\
		1.05 & 15 & -29 & 9 & 324 & 100 & 2 & $1.17 \times 10^{-6}$ & 72.9 \\
		1.05 & 15 & -40 & 9 & 324 & 100 & 2 & $1.20 \times 10^{-6}$ & 72.9 
	\end{tabular}}
\end{minipage}
	\caption[Thresholds relaxed systems considering only lower bands]{Threshold values for the relaxed systems when only the lower two valence bands are considered.}\label{tab:res3}	
\end{table}
\begin{table}[h]
	\begin{minipage}{\linewidth}
	\resizebox{\linewidth}{!}{%
		\begin{tabular}{l|c|c||c|c|c|c||c|r}
			$\theta$ $\left[ ^{\circ}\right]$ & $\Delta_t$ $\left[ meV \right]$ & $\Delta_b$ $\left[ meV \right]$  & $N_{\mathbf{G}}$ & $N_{\mathbf{k}}$ & $N_{\mathbf{R}}$ & $N_{bands}$ & $\tau$ & $p_0$ $ \left[ \% \right]$ \\ \hline
			1.05 & 15 & 0 & 9 & 324 & 100 & 2 & $9.4 \times 10^{-7}$ & 69.4 \\
			1.05 & 15 & -7.9 & 9 & 324 & 100 & 2 & $9.1 \times 10^{-7}$ & 68.6 \\
			1.05 & 15 & -29 & 9 & 324 & 100 & 2 & $1.03 \times 10^{-6}$ & 70.0 \\
			1.05 & 15 & -40 & 9 & 324 & 100 & 2 & $1.15 \times 10^{-6}$ & 71.8 
	\end{tabular}}
	\end{minipage}
	\caption[Thresholds relaxed systems considering only higher bands]{Threshold values for the relaxed lattice when only the higher two conduction bands are considered.}\label{tab:res4}
\end{table}

Let us now present the detailed results coming from the tight-binding calculations, discussing different twist angles in Tabel \ref{tab:res5}. For all cases, we considered four bands around charge neutrality and chose $N_{\mathbf{k}}=900$ where convergence has been checked. In all cases, we obtain irreducibility well above the numerical error. 
\begin{table}[h]
	\begin{minipage}{0.9\linewidth}
	\begin{tabular}{l|c|c||c|r}
		$\theta$ $\left[ ^{\circ}\right]$ & relaxed & $\Delta$ $\left[ meV \right]$ & $\tau_c$ & $p_0$ $\left[ \% \right]$ \\ \hline
		1.05 & no & 0 & $8.32 \times 10^{-5}$ & 89.1 \\
		1.20 & no & 0 & $8.37 \times 10^{-5}$ & 83.4 \\
		1.61 & no & 0 & $3.31 \times 10^{-4}$ & 91.5 \\ \hline
		1.05 & yes & 0 & $1.24 \times 10^{-4}$ & 93.0 \\
		1.20 & yes & 0 & $2.11 \times 10^{-4}$ & 94.6 \\
		1.61 & yes & 0 & $3.84 \times 10^{-4}$ & 93.9 \\ \hline
		1.05 & no & 3 & $8.28\times 10^{-5}$ & 89.1 \\
		1.05 & no & 10 & $7.82 \times 10^{-5}$ & 88.3 \\
		1.05 & yes & 3 & $1.26 \times 10^{-4}$ & 92.8 \\
		1.05 & yes & 10 & $1.19 \times 10^{-4}$ & 92.1
		
	\end{tabular}
	\end{minipage}	
	\caption[Thresholds TBM]{Threshold results for the relaxed and non-relaxed lattice based on the tight-binding  model.}\label{tab:res5}
\end{table}
\FloatBarrier 
\end{document}